\documentclass{moriond}

\def\be{\begin{equation}}
\def\ee{\end{equation}}
\def\bea{\begin{eqnarray}}
\def\eea{\end{eqnarray}}

\newcommand{\Photo}{}

\begin{document}
\vspace*{4cm}
\title{A bonus complementarity in Simplified Models of Dark Matter}

\author{ Bryan Zald\'ivar }

\address{Service de Physique Th\'eorique, Universit\'e Libre de Bruxelles, B-1050 Brussels, Belgium}

\maketitle\abstracts{
Nowadays there is an active discussion about the definition of Simplified Models of Dark Matter (SMDM) as a tool for interpreting LHC searches. Here we point out an additional simplified set-up which captures a very well motivated mechanism beyond the Standard Model: the kinetic-mixing of an extra $U'(1)$ gauge symmetry. In addition to that, even if most of the attention has being paid on LHC ``mono-signals'', here we highlight an unavoidable signature appearing in SMDM with s-channel mediators: dijets or dileptons with no missing energy. We translate these searches into {\it lower} bounds on the DM couplings to the visible sector, showing the nice complementarity with the previous analyses, such that the parameter space of DM is being reduced from above and from below. }


\section{Introduction}

The searches for Dark Matter (DM) have become one of the most actives lines of research at the LHC in the last few years. Departing from the Supersymmetry framework, the Effective Field Theory (EFT) approach provided a very simple, partially model-independent tool where the different signatures at colliders could be interpreted.\footnote{See \cite{Fox:2011fx} and \cite{Buckley:2011vs} for very early works on the subject, and \cite{Mambrini:2011pw} for a complementarity between the two.}

 In particular topologies like monojets \cite{Aad:2015zva} \cite{Khachatryan:2014rra}, monophotons \cite{Aad:2014tda} \cite{Khachatryan:2014rwa}, or monoleptons \cite{ATLAS:2014wra} \cite{Khachatryan:2014tva} (among others) accompanied by large missing transverse energy -which are quoted as ``mono-signals'' in general- have become the main avenues for DM at the LHC. However, soon after the first analyses were done, it became clear the disadvantages of the EFT, whose validity is limited for events with low momentum transfer \cite{Busoni:2013lha}. This fact motivated the community to start thinking about simplified -yet more complete- set-ups, where the extra degrees of freedom are not integrated out, thus rendering the event-by-event analysis valid for all the regions of the parameter space.
 
These Simplified Models of Dark Matter (SMDM) are becoming the next framework within which DM-related searches at the LHC are being interpreted.\footnote{See e.g. refs.\cite{Buchmueller:2014yoa} \cite{Abdallah:2014hon}. See also \cite{Calibbi:2015nha} for a very recent analysis where the DM is studied in a simplified framework, inspired from -but more general than- the neutralino case. }  They consist of very simple (low energy \footnote{In the sense that they do not attempt to be UV completed theories.}) lagrangians with only 3 or 4 parameters: the DM mass $m_\chi$, the mass of the mediator $M_{\rm med}$, and two couplings, DM-mediator $g_\chi$ and mediator-SM $g_q$, in the case of ``portals'' (i.e. s-channel processes), or one coupling $g_t$ for the DM-mediator-SM interactions, if considering t-channel processes. Even if popularly used in the context of WIMP searches, other DM alternatives are also analysed in this framework for LHC studies \cite{Daci:2015hca}.

Although in SMDM implementations it has being usually assumed that the coupling $g_q$ is universal for all quarks -thus simplifying enormously the interpretation of the experimental results-,  a priori this may not be the case for the low energy predictions of more fundamental models. Thus it is interesting to ask the question of which are the ``ultraviolet'' completions of such SMDM realisations and, at the same time, how can such implementations be extended in such a way as to represent other possible theoretically well motivated scenarios. This should be achieved without spoiling their minimality; specifically, without enlarging the number of free parameters which render them feasible for LHC interpretations. 

Another aspect concerns the possible signatures at the LHC. The introduction of an explicit mediator between the dark matter and the SM implies that not only processes like SM+SM$\to$mediator$\to$DM+DM are possible (leading to monojets$+E^T_{\rm miss}$ for example), but inevitably, the possibility SM+SM$\to$mediator$\to$SM+SM should also be taken into account (leading to dileptons or dijets with no $E^T_{\rm miss}$). We quantify the complementarity existing between these two kinds of processes, together with the constraints coming from dark matter Direct Detection searches. As we will see, the two strategies constrain the DM parameter space from opposite directions, efficiently excluding large parts of the parameter space of such SMDM set-ups.
\newline\newline
In the next sections we will first briefly revise the theoretical motivations for some of the existing SMDM set-ups; specifically focusing in the s-channel configurations and vector mediators. We identify possible alternatives capturing other interesting textures which may be connected to popular UV completions. We then dedicate the rest of the presentation to describe how the complementarity between mono-signals and disignals is achieved.

\section{From the UV to Simplified Models}
Here we adopt a top-down approach, going from the generic elements of popular complete models to the low energy parametrisations defined by SMDM. We concentrate in ``s-channel'' set-ups, i.e. those where the dark matter does not have direct interactions with the SM particles but only indirectly via a ``portal'' which couple to both the dark and the visible sectors. Here we restrict to the case of vector mediators. 
\newline\newline\noindent
Typically the introduction of new vector particles is associated with the postulation of new gauge symmetries. The simplest of these is an abelian $U'(1)$, which is one of the best motivated extensions of the Standard Model. The generic lagrangian for the interaction of a $Z'$ with fermions is:
\be
{\cal L}_{\rm NC} = g_X J^\mu_{\rm NC} Z'_\mu,~~~~ J^\mu_{\rm NC} = \sum_i \bar f_i \gamma^\mu [g^i_V - g^i_A\gamma^5]f_i,
~~~~g^i_{V,A} = \frac{Q^i_L \pm Q^i_R}{2}~;
\label{NC}
\ee
where $J^\mu_{\rm NC}$ is the associated neutral current, containing every possible fermion $f_i$. The couplings $g_V$, $g_A$ are the vector and axial couplings respectively, where $Q^i_{L,R}$ are the left and right charges of every fermion $i$ to $Z'$. Note from (\ref{NC}) that  the gauge coupling $g_X$ is not really a parameter independent of the charges, or viceversa. Let the DM, $\chi$, be on of the above fermions, such that its couplings with the $Z'$ are in general $g^\chi_V, g^\chi_A$. To generate the $Z'$ and DM masses can be done in several ways, so one could simply take as working variables the masses themselves $m_\chi, m_{Z'}$, independently on the way they were obtained from the fundamental parameters. From the above arguments we see that we can work in a parametrisation with the following independent variables:
\be
m_{Z'}, m_\chi, g^i_V, g^i_A, g^\chi_V, g^\chi_A ~.
\label{dof}
\ee
where $i$ represent every possible SM fermion \footnote{So a priori there are 21 or these $g^i$'s, where RH neutrinos are not counted. However, the requirement for anomaly cancellations may reduce this number, since they constrain the charges to the $Z'$.}. All in all there are, a priori, 25 parameters in expr.(\ref{dof}). Most of the (complete) models that can be written regarding a $Z'$ will present the above ingredients. They will differ in the assumptions for the couplings, but also, in the sector responsible for giving mass to the $Z'$ (like the number of extra scalars acquiring VEVs, or how these scalars are charged under the $U'(1)$ or the SM gauge group). Also, a complete model should take care of the anomalies that are a priori naturally present whenever we have a new gauge sector. 

Next we mention some of the most popular models involving $Z'$ bosons \cite{Langacker:2008yv}:\newline\newline
\underline{Sequential Model.} In this set up the $Z'$  couplings to the SM fermions are fixed to be equal to those of the $Z$ boson itself. This is just a ``reference'' realisation, since it says nothing about anomaly cancellations or the origin of the $Z'$ mass. If considered together with DM, this model has only 4 independent, unknown parameters (the two couplings of the DM and the $m_\chi, M_{Z'}$ masses). 
\newline\noindent
\underline{$B-L$ Model.} By gauging the $B-L$ quantum number,  the resulting model is anomaly free with just the matter content of the SM plus RH neutrinos.\footnote{See \cite{Crivellin:2015mga} for a recent alternative where a different combination of lepton number is gauged, which is motivated by the recent LHCb anomalies  \cite{Aaij:2013qta}.} The associated $Z'$ will have only pure-vector couplings to the SM fermions (i.e. $g^i_A=0$ in this model).\footnote{Except for the neutrinos if they are Majorana.} This is one of the simplest $Z'$ realisations. It contains an extra free parameter (w.r.t. the Sequential Model), which may be thought as e.g. the new gauge coupling $g_{BL}$. We have then 5 independent parameters, since the different $g^i_V$ are determined by anomaly cancellation (modulo this overall multiplicative $g_{BL}$ factor).
\newline\noindent
\underline{$E_6$ models.} This is a realisation motivated from string theory, which is anomaly free thanks to exotic fields that may be decoupled from the low energy phenomenology.  It actually consist of two $U'(1)$'s, usually called $U(1)_\chi$ and $U(1)_\psi$, with the corresponding charges of all the fermions to the two gauge bosons. Most studies assume that only one $Z'$, coupling  to the linear combination $Q=\cos\theta_{E_6} Q_\chi + \sin\theta_{E_6} Q_\psi$, is relevant at low energies. A priori there are then 3 extra parameters: $\theta_{E_6}, g_\chi, g_\psi$. In one of the realisations, $\theta_{E_6}=\pi/2$, such that only $Z_\psi$ is relevant. Actually in this case all the couplings of the SM fermions to $Z_\psi$ turn out to be purely axial (i.e. $g^i_V=0$). This latter realisation has, as the previous case, 5 independent parameters; i.e. $g_\psi, M_{Z'}, m_\chi, g^\chi_V$ and $g^\chi_A$. 
\newline\newline\noindent
 \underline{Kinetic Mixing.}  This is one of the favourite implementations of an extra $U'(1)$, since many of the existing New Physics models in the literature lead to this kind of set up. Here it is worth commenting a little bit more in detail. For example, Salvioni {\it et al} \cite{Salvioni:2009mt} have presented a minimal model which is a combination of $B-L$ and hypercharge $U'(1)$, the latter being directly realised by the Kinetic Mixing (KM) set-up. There it is shown how different linear combinations of $B-L$ and KM can generate a whole family of models which predict the existence of $Z'$ particles. In fact, this set up is by now used by the ATLAS collaboration itself \cite{Aad:2014cka}.

In KM the observation is that a kinetic term like $-\frac{\epsilon}{2}B_{\mu\nu} X^{\mu\nu}$ is actually gauge invariant. Here $\epsilon$ is just the kinetic mixing coupling, $B_{\mu\nu}$ the SM stress tensor for $U(1)_Y$, and $X_{\mu\nu}$ the stress tensor of the new gauge group $U'(1)$. Even if at tree level $\epsilon=0$, it may be generated at loop level if there are particules in the theory coupled to both gauge groups. After redefinition of the gauge fields to render the kinetic terms diagonal, the new fields get mixed once the electroweak symmetry is broken. The resulting mass eigenstates are thus:
\bea
Z'_\mu &=& (\cos\theta_W W_\mu^3 - \sin\theta_W B_\mu)\sin\alpha + X_\mu\cos\alpha ~\nonumber\\
Z_\mu &=& (\cos\theta_W W_\mu^3 - \sin\theta_W B_\mu)\cos\alpha - X_\mu\sin\alpha ~,
\eea
where $\theta_W$ is the Weinberg angle and $\alpha$ the mixing, where $\alpha=0$ for $\epsilon=0$. The SM fermions, even if not being charged w.r.t. the $U'(1)$, will thus couple to the physical $Z'$ as:
\be
\bar f_i\gamma^\mu [(g_{V,\rm SM}^i - g_{A,\rm SM}^i \gamma^5)\sin\alpha] f_i Z'_\mu
\ee
where $g^i_{(V,A),\rm SM}$ are the standard couplings of the SM fermions to the $Z^{\rm SM}_\mu$ boson. Similarly, even if the DM is not coupled to the SM $U(1)_Y$, the kinetic mixing will induce a coupling of DM to the $Z_\mu$ as:
\be
\bar \chi \gamma^\mu (g_\chi\sin\alpha) \chi Z_\mu~,
\ee
where we have assumed that DM couples only with vector-like coupling to the $U'(1)$. The 4 independent parameters of this set-up can be taken to be: the $Z'-Z$ mixing angle $\sin\alpha$, the DM coupling $g_\chi$, and the two masses $m_{Z'}$ and $m_\chi$. 

\subsection{Towards simplified models}
Above we have seen some examples of models where the couplings can be purely vectorial (as for $U(1)_{BL}$), purely axial (as $U(1)_\psi$ in $E_6$) or a mix of both vector and axial, as the Kinetic Mixing set-up. In the $U(1)_{BL}$ case for example, $u$ and $d$ quarks actually have identical couplings, and the leptons have also identical couplings among themselves. In the $U(1)_\psi$ case, quarks and leptons have all identical axial couplings in fact. The couplings to the 2nd and 3rd fermion generations are identical as the 1st one in all of these cases. \footnote{See \cite{Calibbi:2015sfa} for a recent dark matter model whose $Z'$ portal not only have non-universal couplings, but which also allows for flavour-violating couplings. } 

Now we make the connection with the simplified models that has been lately considered in the literature. We see that \cite{Buchmueller:2014yoa} \cite{Abdallah:2014hon}: \footnote{Note that couplings to quark are universal in these set-ups.}
\be
-{\cal L}_V \equiv g_q\sum_q  Z'_\mu \bar q\gamma^\mu q + g_\chi Z'_\mu\bar\chi\gamma^\mu\chi~,~~~
-{\cal L}_A \equiv g_q\sum_q  Z'_\mu \bar q\gamma^\mu\gamma^5 q + g_\chi Z'_\mu\bar\chi\gamma^\mu\gamma^5\chi~,
\label{smdm}
\ee
actually correspond to some of the above realisations. Indeed ${\cal L}_V$ may be motivated by a $U(1)_{BL}$-type model, where $g_q$ parametrizes the overall multiplicative factor to all quarks. If sticking to this motivation, it should be noted that if we want to extend the ${\cal L}_V$ model to the lepton sector, a priori the couplings there would be different w.r.t. the quarks (but identical among themselves). In the same fashion, ${\cal L}_A$ may be motivated by some $E_6$ realisations  \footnote{However, contrary to the $U(1)_{BL}$ case where the breaking scale can be a priori any, in the $E_6$ case one expects the scale to be very high, order $E_{\rm GUT}$ or lager, since this scenario is motivated by string theory.}
, $U(1)_\psi$ above, where we have only axial couplings. Thus, for ${\cal L}_A$, the lepton sector may have identical couplings as the quarks.
On the other hand, none of the above simplified models capture the textures of the KM set-up, where the $u_L, u_R,d_L, d_R$ quarks all have different $V,A$ couplings. 
\newline\newline
It is then very well motivated to look for a SMDM which capture the KM realisation. To recap, there are only 4 unknown parameters: $m_\chi, m_{Z'}, \sin\alpha, g_\chi$, such that the couplings are:
\bea
g^i_{(V,A)\rm SM}\sin\alpha&:&~~~~~Z'~{\rm couplings~to~fermions} \\
g^i_{(V,A)\rm SM}\cos\alpha&:&~~~~~Z~{\rm couplings~to~fermions}\nonumber \\
g_\chi\cos\alpha&:&~~~~~Z'~{\rm couplings~to~DM}\nonumber\\
g_\chi\sin\alpha&:&~~~~~Z~{\rm couplings~to~DM}~.\nonumber
\eea
Adopting such a simple set-up in addition to the other ones already being considered (\ref{smdm}) would definitely spam a large class of scenarios beyond SM where a new $Z'$ interacts with the SM via a kinetic-mixing mechanism.

\section{An LHC complementarity for SMDM}
Here we focus in an s-channel, vector mediator set-up as the one shown in (\ref{smdm}). However the following discussion is equally valid for models with scalar mediators. 
It is evident that such model will not only contribute to, for example, monojet signals plus missing energy, but also to signatures with two energetic jets or leptons in the final state, without missing energy. Specifically, the searches for resonances in the dilepton \cite{Aad:2012hf} \cite{Chatrchyan:2012oaa} or  dijet  \cite{ATLAS:2012pu} \cite{Chatrchyan:2013qha} distributions are among the strongest bounds to New Physics. 

\begin{figure}[htbp]
\centering\includegraphics[width=0.49\textwidth]{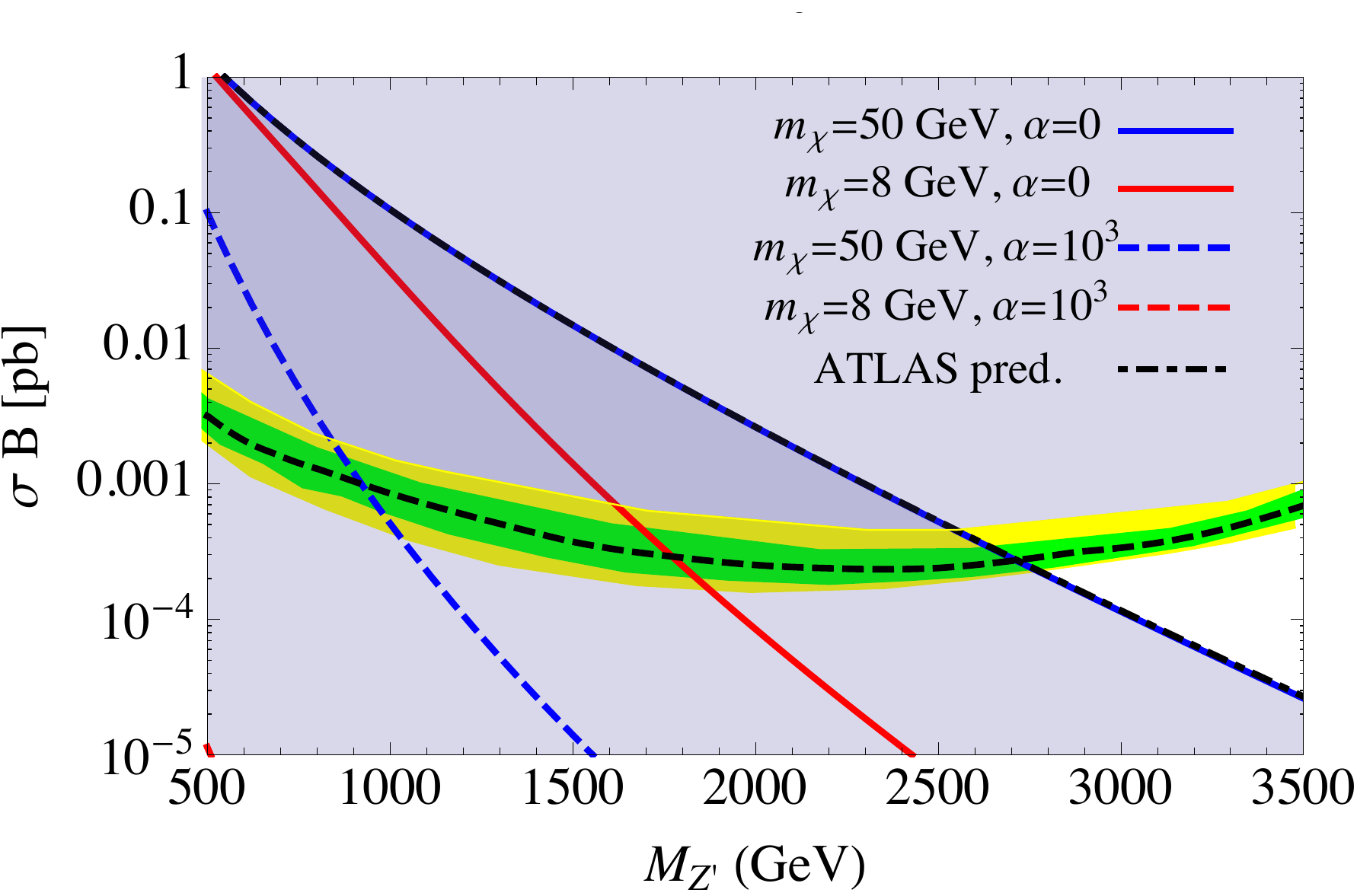}
\centering\includegraphics[width=0.49\textwidth]{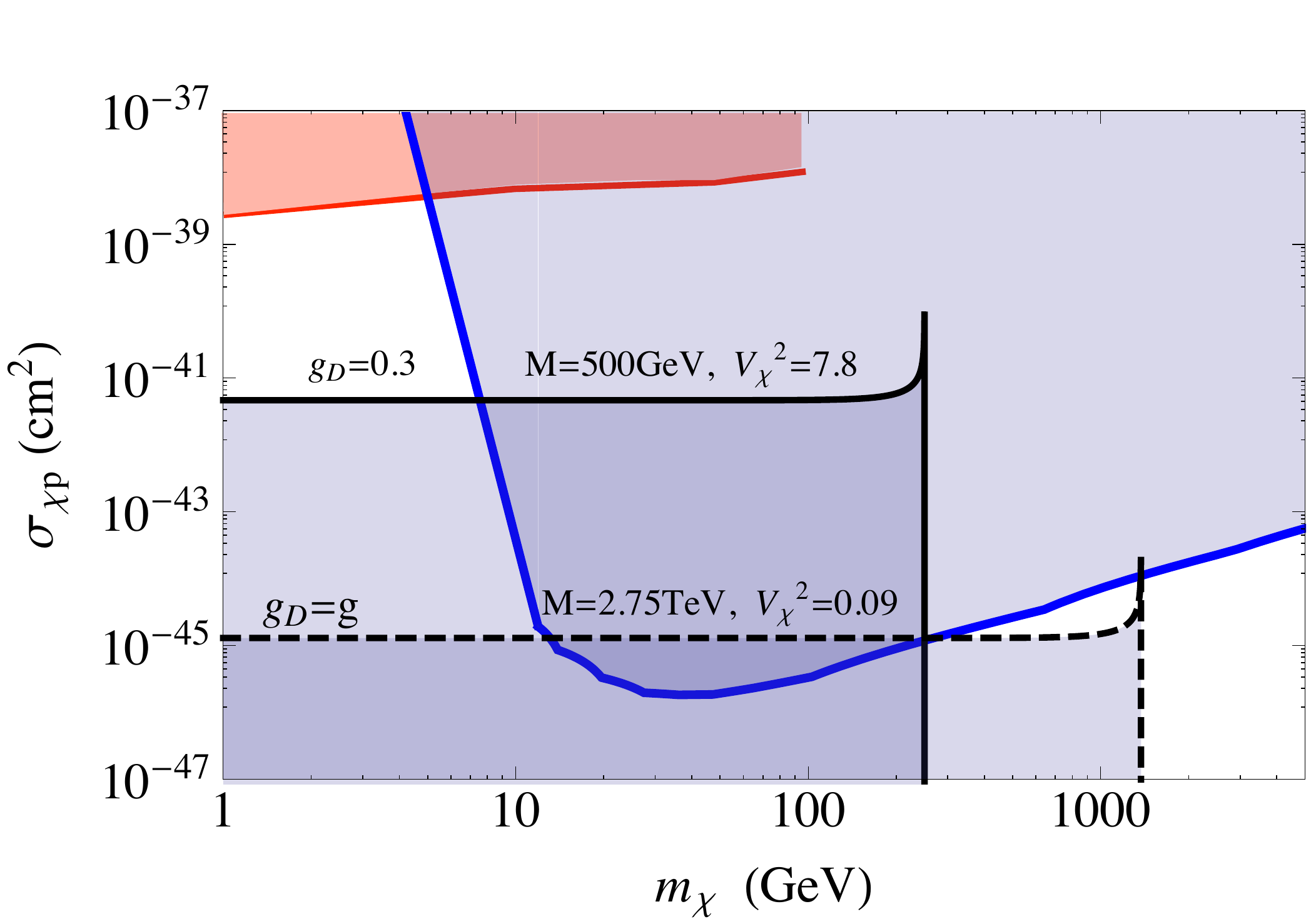}
  \caption{Left) Dilepton production cross section at the LHC, and limits from resonance searches by ATLAS (brazilian band). The different lines show the bounds on the model for different choices of parameters, according to the LUX experiment. Right) Spin-independent cross section for DM-nucleon scattering. The LUX bound is shown in solid blue, and the monojet bound is shown in red. The black lines correspond to bounds coming from dileptons, for different choices of parameters. In all cases the shaded regions are excluded.}
  \label{plots}
\end{figure}

The key observation here is that the existence of an invisible $Z'$ branching ratio (provided it can decay to DM) weakens the current LHC limits \cite{Arcadi:2013qia}. Indeed at the partonic level the cross section for $\sigma(pp\to Z'\to \bar ff)$, having a Breit-Wigner profile:
\be
\sigma(pp\to Z'\to \bar ff)\approx \frac{1}{12\pi}(|g^q_V|^2 + |g^q_A|^2)(|g^f_V|^2 + |g^f_A|^2)\frac{s}{(s-m^2_{Z'})^2 + \Gamma^2_{Z'}m^2_{Z'}}
\ee 
can be re-expressed as:
\bea
\sigma(pp\to Z'\to \bar ff)&\approx& \frac{1}{12}(|g^q_V|^2 + |g^q_A|^2)(|g^f_V|^2 + |g^f_A|^2)\nonumber\\
&\times&\frac{m_{Z'}}{\Gamma^{v}_{Z'}}(1-Br_\chi) \delta(s-m^2_{Z'})
\eea
after approximating the Breit-Wigner by a Dirac delta. Here $Br_\chi$ is the branching ratio of $Z'$ to DM and $\Gamma^{v}_{Z'}$ the total width of $Z'$ to the visible sector. Thus, for given masses and couplings of $Z'$ to the visible sector, the dilepton (or dijet) production cross section diminishes when $Br_\chi$ increases.
\newline\newline
However, the invisible branching ratio cannot increase arbitrarily. Upper bounds on it are imposed by the LHC itself with mono-signal searches, but also, by Direction Detection, which in the case of spin-independent (SI) DM-nucleon scatterings are particularly strong. Given a value for the SI cross section $\sigma^{\rm SI}_{\chi N}$, for example, taken from the experimental limit for a given DM mass, one can extract un upper bound on the $Br_\chi$ as:
\be
Br_\chi = \left[1 + \left(\frac{2\mu_{\chi N}}{m^2_{Z'}}\right)^2 \frac{\tilde c_F\alpha^{\rm SI}_{Z,A}}{\pi(1+\alpha^2)\sigma^{\rm SI, exp}_{\chi N}}\right]^{-1}
\label{BRx}
\ee  
where $\alpha^{\rm SI}_{Z,A}\equiv[g_V^u(1+Z/A) + g_V^d(2-Z/A)]^2$ is a function of the charge number $Z$ and mass number $A$ of the nucleus that contains the nucleon; $\mu_{\chi N}$ is the DM-nucleon reduced mass; $\alpha\equiv g_A^\chi/g_V^\chi$ is the ratio of DM axial to vector coupling to $Z'$ and $\tilde c_F\equiv\sum_f c_f (|g^f_V|^2 + |g^f_A|^2)$. Of course the experimental limit $\sigma^{\rm SI, exp}_{\chi N}$ will depend on the DM mass. For example, by applying (\ref{BRx}) we find that for $m_{Z'}\sim3$ TeV and pure vector couplings ($\alpha=0$), the $Br_\chi$ can be even 90\% if $m_\chi\sim 6$ GeV, but for $m_\chi\sim40$ GeV which is around the point of maximum sensitivity for LUX \cite{Akerib:2013tjd}, $Br_\chi$ is constrained to be below 1\%.  
\newline\newline
For that reason it is motivated to combine the two analysis: by how much we can relax the bounds on $m_{Z'}$ from dijets and dileptons, will depend on LUX, if we allow the $Z'$ to couple to the DM. In fig.\ref{plots}-left) we show the specific case of the interplay between LUX and the dilepton bounds coming from the LHC.\footnote{See \cite{Alves:2013tqa} for an alternative study on dijets instead.} By assuming a Sequential Model (see previous section), the ATLAS model prediction \cite{Aad:2012hf}  (which do not assume any coupling to DM) is shown by the black dot-dashed line. The brazilian exclusion band thus forbids $Z'$ masses below 2.7 TeV or so. Now by adding a coupling to DM (i.e. an invisible branching fraction) such a bound can even go down to $\sim$1 TeV, even for DM mass of 50 GeV which and couples mainly axially to the $Z'$. This is shown by the blue dashed line. Of course if the coupling is mostly vector-like, the LUX searches are much more sensitive even for a lighter mass, and the dilepton bound can only get down to 1.7 TeV or so (solid red line). However, at $m_\chi=50$ GeV the LUX bound is so strong, that for pure-vector couplings a $Z'$ of $\sim2.7$ TeV or less is not allowed to have an appreciable amount of invisible branching ratio, so the ATLAS bound remain unchanged (solid blue line).

An even more interesting way to see the same effect in shown in fig.\ref{plots}-right. There we put the dilepton searches in the Direct Detection plane, where we compare directly with LUX (blue line)  together with the monojet bounds (red). The solid black line corresponds to the exclusion by dilepton, for $m_{Z'}=500$ GeV and a a dark coupling $g^\chi_V=0.3\cdot\sqrt{7.8}=0.83$. The lower bound comes because for this $Z'$ mass, a smaller $Br_\chi$ (leading to a smaller $\sigma^{\rm SI}_{\chi N}$) would render the $Z'$ {\it too visible} according to the dilepton searches. Of course, for $m_\chi>m_{Z'}/2$ there is no interplay with the dilepton constraint since $Br_\chi = 0$. Also, for a larger $Z'$ mass (as the one in black dashed line) the dilepton constraints are less severe, so the lower limit for $\sigma^{\rm SI}_{\chi N}$ is weaker. See more details at \cite{Arcadi:2013qia}.

In any case, it is very interesting to observe how while the monojet and direct detection searches are excluding the parameter space of the model from above, the dilepton searches are excluding it from below. The same remains true for other mono-signals, as compared to dijets, instead of dileptons. The qualitative features of this analysis are also independent of the nature of the mediator, as long as it leads to an s-channel DM-SM portal.

\section{Conclusions}
We have tried to motivate the implementation of a Simplified Model of Dark Matter which, while being equally ``economical'' (concerning the number of independent parameters) with respect to the more popular implementations, is able to represent a large class of theoretical scenarios beyond Standard Model which contain an extra $U'(1)$ gauge symmetry, where the corresponding gauge boson kinetically mixes with the SM $Z$ boson. 

A posteriori, we have shown that any Simplified Model where the DM communicates to the SM via an s-channel mediator (regardless its nature), will have consequences not only on mono-signal searches, but also di-signal searches, both of them complementing  each other. We have illustrated this fact in a particular example where we compare monojets and dileptons searches, together with Direct Detection bounds, in a model where the $Z'$ couples to the Standard Model \`a la Sequential Model. 

This sort of complementarity could be very useful at Run II of the LHC, where the searches for dark matter will play a major role.

\section*{Acknowledgments}

I thank the organisers of the {\it Moriond EW 2015} for financial support and the opportunity to present my results, and J. Heeck for useful discussions.

\end{document}